\documentclass[preprint,review,12pt]{elsarticle}

\usepackage{amssymb}
\usepackage{amsmath}
\usepackage{amsthm}
\usepackage{nicematrix}

\usepackage{graphicx}
\usepackage{graphics}
\usepackage{rotating}
\usepackage{subfig}
\usepackage{tikz}
\usepackage[ruled,vlined,linesnumbered]{algorithm2e} 
\usepackage{lscape}
\usepackage{multirow}
\usepackage{hyperref}
\usepackage{enumitem}
\usepackage{array}
\usepackage[misc,geometry]{ifsym}
\usepackage{booktabs}

\theoremstyle{remark}
\newtheorem*{remark}{Remark}

\theoremstyle{example}
\newtheorem*{example}{Example}

\usetikzlibrary{arrows.meta}

\def\tu#1{\ensuremath{\langle #1 \rangle}}

\def\tu#1{\langle #1\rangle}
\newcommand{\A}{{\bf A}}
\newcommand{\I}{{\bf I}}
\newcommand{\B}{{\bf B}}

\journal{Information Sciences}

\begin{document}
\begin{frontmatter}



\title{GreCon3: Mitigating High Resource Utilization of GreCon Algorithms for Boolean Matrix Factorization}


\author{Petr Kraj\v{c}a\corref{cor1}} 
\ead{petr.krajca@upol.cz}

\author{Martin Trnecka} 
\ead{martin.trnecka@upol.cz} 
\address{Department of Computer Science, Palack\'y University Olomouc, Czech Republic}
\cortext[cor1]{corresponding author}



\begin{abstract}
Boolean matrix factorization (BMF) is a fundamental tool for analyzing binary data and discovering latent information hidden in the data. Formal Concept Analysis (FCA) provides us with an essential insight into BMF and the design of algorithms. Due to FCA, we have the GreCon and GreCon2 algorithms providing high-quality factorizations at the cost of high memory consumption and long running times. In this paper, we introduce GreCon3, a substantial revision of these algorithms, significantly improving both computational efficiency and memory usage. These improvements are achieved with a novel space-efficient data structure that tracks unprocessed data. Further, a novel strategy incrementally initializing this data structure is proposed. This strategy reduces memory consumption and omits data irrelevant to the remainder of the computation. Moreover, we show that the first factors can be discovered with less effort. Since the first factors tend to describe large portions of the data, this optimization, along with others, significantly contributes to the overall improvement of the algorithm's performance. An experimental evaluation shows that GreCon3 substantially outperforms its predecessor GreCon2. The proposed algorithm thus advances the state of the art in BMF based on FCA and enables efficient factorization of datasets previously infeasible for the GreCon algorithm.
\end{abstract}




\end{frontmatter}



\newcommand{\pk}[1]{\textcolor{red}{#1}}

\newcommand{\mt}[1]{\textcolor{blue}{#1}}

\section{Introduction}

Boolean matrix factorization (BMF) is a well-established and widely used method of data analysis~\cite{recent}, which aims to decompose an input binary matrix into two smaller binary matrices that retain the information of the original one, thereby providing an explanation of the data in terms of latent factors. Belohlavek and Vychodil, in their pioneering work~\cite{BeVy10:ICSS}, established a link between BMF and Formal concept analysis (FCA, \cite{Ganter2024}), which provides a fundamental framework for the analysis of binary data. Based on this connection, they proposed two algorithms, GreCon and GreConD. Both algorithms are considered a baseline in BMF nowadays. While the GreConD algorithm is very fast, GreCon is significantly slower, as it requires several iterations over a potentially large set of so-called formal concepts, which serve as candidates for factors; however, it produces better results.

Trnecka and Vyjidacek~\cite{TrVy22:KBS} revised the GreCon algorithm and proposed its improved version, called GreCon2. The new version preserves the results of GreCon while outperforming the original algorithm in terms of running time.\footnote{Note that the number ``2'' in the algorithm’s name reflects its version. The results obtained using GreCon and GreCon2 are identical.} In some cases, GreCon2 even outperforms the GreConD algorithm. Moreover, the authors show that the search spaces of GreConD and GreCon differ significantly. GreConD considers only a small fraction of formal concepts, whereas GreCon explores all of them. As a consequence, GreCon may discover better factors that remain unconsidered by GreConD.

Although the GreCon2 algorithm provides a significant speed improvement over GreCon and performs well on reasonably sized (preferably smaller) data, we have identified multiple spots where GreCon2 can be improved to obtain an even faster algorithm capable of efficiently handling larger matrices. One of the weakest spots of GreCon2 is the data structure that keeps the necessary information for factor computation. Its role is to suppress the recomputation of coverage for all formal concepts, which is the most time-consuming task in the original GreCon. A drawback of this structure, not mentioned in~\cite{TrVy22:KBS}, is that it requires a considerable amount of memory, which limits the algorithm when applied to large data. Moreover, GreCon2 spends a substantial amount of time initializing this data structure.


The main contribution of this paper is a novel BMF algorithm called GreCon3 representing a revised version of GreCon and GreCon2. In the third version of the GreCon algorithm, we introduce a more memory-efficient data structure designed to store and update information about the coverage of formal concepts during factor computation. Furthermore, we propose an alternative, less time-consuming strategy for its initialization. Additionally, we present several further optimizations that result in a substantial improvement in both memory usage and runtime. Finally, we provide an experimental evaluation showing that GreCon3 shifted the GreCon algorithm to a further level, as it significantly outperforms GreCon2.

The rest of the paper is organized as follows. In Section~\ref{sec:preliminaries}, we provide the basic notions and notations used throughout the paper. We also describe the connection between BMF and FCA. Section~\ref{sec:algorithm} presents a detailed description of the GreCon3 algorithm together with the motivation behind all proposed optimizations. Section~\ref{sec:evaluation} summarizes the experimental comparison of GreCon3, GreCon2, and GreConD, and provides several implementation notes. Finally, Section~\ref{sec:conclusion} concludes the paper and outlines directions for future research.

\section{Preliminaries}
\label{sec:preliminaries}


Boolean matrix factorization and related problems have been investigated since the 1970s, when it was shown that the fundamental problem is NP-hard~\cite{Stockmeyer75}. Today, BMF is employed in data mining~\cite{tiling,dbp,hyper}, machine learning~\cite{hess,ravan}, knowledge management~\cite{panda,sadok}, and numerous other fields~\cite{graf,role,bio}.

The paper focuses on a from-below Boolean matrix factorization~\cite{ess}, i.e., the main problem is for Boolean matrix $\I\in\{0,1\}^{m\times n}$ with $m$ rows and $n$ columns to find factor matrices $\A\in\{0,1\}^{m\times k}$ and $\B\in\{0,1\}^{k\times n}$ for which $(\A \circ \B) \leq \I$ where $\circ$ denotes the Boolean matrix multiplication, i.e., $(\A\circ \B)_{ij} = \max_{l=1}^k 
(\min(\A_{il},\B_{lj}))$. Quality of approximate decomposition is expressed as a metric $E(\I, \A \circ \B)$ representing a number of entries of $\I_{ij} = 1$ for which $(\A\circ \B)_{ij} = 0$, i.e., $E(\I, \A \circ \B) = || \I - (\A \circ \B) ||$.\footnote{Note that $||\cdot||$ represents the $L_{1}$ norm.}


We are interested in matrices where $k$ and the value of metric $E(\I, \A \circ \B)$ is reasonably small. In other words we are looking for matrices $\mathbf A$ and $\mathbf B$ that contain reasonable number of factors and explain reasonable large part of the data.

A factorization (or also decomposition) of $\I$ into $\A \circ \B$ is interpreted as a discovery of $k$ factors that exactly or approximately describe object-attribute data $\I$ via  object-factor matrix $\A$ and factor-attribute matrix $\B$.

Factors are latent variables hidden in the data. A factor is described by the objects to which it applies and by the attributes that represent its manifestation. In general, a factor can be viewed as a rectangle in the data matrix $\I$ (after a suitable permutation of rows and columns). Matrices $\mathbf{A}$ and $\mathbf{B}$ capture the factors: the $l$-th factor is described by the $l$-th column of $\mathbf{A}$ and the $l$-th row of $\mathbf{B}$. A simple example follows.

\begin{example}
Let us considered the following Boolean matrix
\[
\mathbf{I} =
\begin{bmatrix}
1 & 1 & 1 & 0 & 0 & 0 \\
1 & 1 & 1 & 0 & 0 & 0 \\
0 & 1 & 1 & 1 & 1 & 0 \\
0 & 1 & 1 & 1 & 1 & 1 \\
0 & 0 & 1 & 1 & 0 & 1
\end{bmatrix}.
\]

\noindent This matrix can be decomposed into three factors captured by matrices $\mathbf{A}$ and $\mathbf{B}$ as follows:
\[
\mathbf{A} =
\begin{bmatrix}
1 & 0 & 0 \\
1 & 0 & 0 \\
0 & 1 & 0 \\
0 & 1 & 1 \\
0 & 0 & 1
\end{bmatrix},
\quad
\mathbf{B} =
\begin{bmatrix}
1 & 1 & 1 & 0 & 0 & 0 \\
0 & 1 & 1 & 1 & 1 & 0 \\
0 & 0 & 1 & 1 & 0 & 1
\end{bmatrix}.
\]

\noindent Each factor forms a rectangle in data, i.e.,

\[
\mathbf{I} = \A \circ \B =
\begin{bmatrix}
1 & 1 & 1 & 0 & 0 & 0 \\
1 & 1 & 1 & 0 & 0 & 0 \\
0 & 0 & 0 & 0 & 0 & 0 \\
0 & 0 & 0 & 0 & 0 & 0 \\
0 & 0 & 0 & 0 & 0 & 0
\end{bmatrix}
\vee
\begin{bmatrix}
0 & 0 & 0 & 0 & 0 & 0 \\
0 & 0 & 0 & 0 & 0 & 0 \\
0 & 1 & 1 & 1 & 1 & 0 \\
0 & 1 & 1 & 1 & 1 & 0 \\
0 & 0 & 0 & 0 & 0 & 0
\end{bmatrix}
\vee
\begin{bmatrix}
0 & 0 & 0 & 0 & 0 & 0 \\
0 & 0 & 0 & 0 & 0 & 0 \\
0 & 0 & 0 & 0 & 0 & 0 \\
0 & 0 & 1 & 1 & 0 & 1 \\
0 & 0 & 1 & 1 & 0 & 1
\end{bmatrix},
\]

\noindent where $\vee$ is superposition operation (or logical disjunction).
\end{example}
 

In the paper, we use a tight connection between BMF and Formal concept analysis~\cite{BeVy10:ICSS}. The basic data structure in formal concept analysis is a {\it formal context} which is a triple $\left<{X}, {Y}, {I} \right>$, where ${X}$ is a set of $m$ objects, ${Y}$ is a set of $n$ attributes, and ${I}$ is an incidence relation between ${X}$ and ${Y}$. There is a one-to-one correspondence between formal contexts and Boolean matrices, i.e., each Boolean matrix $\I\in\{0,1\}^{m\times n}$ can be seen as a formal context where $\tu{i,j} \in {I}$ if the entry $\I_{ij} = 1$ and vice versa.


For the sake of simplicity, we use a mix of matrix and set notation, as it is more convenient. To every $\I\in \{0,1\}^{n\times m}$ one may associate a pair
$\tu{{}^{\uparrow}, {}^{\downarrow}}$ of
concept forming operators assigning to sets 
${C} \subseteq {X}=\{1,\dots,m\}$ and ${D}\subseteq  {Y}=\{1,\dots,n\}$ the sets $C^\uparrow\subseteq {Y}$ and $D^\downarrow\subseteq {X}$ defined by
\begin{eqnarray*}
  C^\uparrow = \{ j\in {Y} \,|\, \forall i\in C: \I_{ij}=1\},\\
  D^\downarrow = \{ i\in {X} \,|\, \forall j\in D: \I_{ij}=1\}.
\end{eqnarray*}

A pair $\tu{C,D}$ for which $C^\uparrow = D$ and $D^\downarrow = C$ is called a \emph{formal concept}. The set of all formal concepts for the matrix $\I$ is defined as follows
\[
    {\cal B}(\I) = \{ \tu{C,D} \mid C\subseteq {X}, D\subseteq {Y},
     C^\uparrow=D, D^\downarrow=C\}.
\]

Every set
${\cal F} = \{\tu{C_1,D_1},\dots,\tu{C_k,D_k}\} \subseteq {\cal B}(\I)$, with a fixed indexing of the formal concepts $\tu{C_l,D_l}$
induces the $m\times k$ and $k\times n$
Boolean matrices $\A_{\cal F}$ and $\B_{\cal F}$ by
\begin{eqnarray*}\label{eqn:A}
   (\A_{{\cal F}})_{il}=\left\{
   \begin{array}{l}
     1, \textrm{if}~ i\in C_l,\\
     0, \textrm{if}~ i\not\in C_l,
   \end{array}
   \right.
\end{eqnarray*}
and
\begin{eqnarray*}\label{eqn:B}
   (\B_{\cal F})_{lj}=\left\{
   \begin{array}{l}
     1, \textrm{if}~ j\in D_l,\\
     0, \textrm{if}~ j\not\in D_l,
   \end{array}
   \right.
\end{eqnarray*}
for $l=1,\dots,k$.  The set ${\cal F}$ is called a set of {\it factor concepts}. We say that the entry $\I_{ij} =1 $ is {\it covered} by formal concept $\tu{A,B}$ if $i \in A$ and $j \in B$.

\section{Algorithm}
\label{sec:algorithm}

The original GreCon algorithm~\cite{BeVy10:ICSS}\footnote{In the paper GreCon is called Algorithm~1.} uses a straightforward greedy strategy to identify factor concepts:
\begin{enumerate}
\item It enumerates all formal concepts.
\item For each formal concept, it computes the number of ones it covers.
\item Formal concept covering the largest number of ones is selected as a factor concept and ones that are covered by this formal concept are turned into zeros.
\item Algorithm proceeds with the step 2 until all ones are covered by some factor concept.
\end{enumerate}

Even though the algorithm provides a high quality factorization, steps~1 and~2 are its main weaknesses. Namely, computing of all formal concepts can be very time demanding task. Moreover, there may be so many formal concepts they do not fit into the main memory of computer. To address this issue~\cite{BeVy10:ICSS} proposed the GreConD\footnote{In the paper GreConD is called Algorithm~2.} algorithm which computes formal concepts ``on-demand'' while looking for factor concepts. Since GreConD avoids computing of substantial number of formal concepts, it tends to produce worse factorization than GreCon. 

Few algorithms aiming to decrease the high resource requirements of GreCon have been proposed. For instance, to deal with large amounts of formal concepts \cite{KrVy11:ICDM} proposes to use GreCon along with a fast algorithm for computing frequent closed itemsets (LCMv3~\cite{lcm3}) to find an initial approximate decomposition. Then, GreConD is used to compute the remaining factor concepts. Since the set of frequent closed itemsets is a subset of formal concepts, this algorithm tends to provide results that are not as good as those provided by GreCon. In fact, this algorithm trades an increase in computation speed for a~decrease in result quality.

In GreCon2~\cite{TrVy22:KBS}, Trnecka and Vyjidacek address the performance bottleneck related to computing coverage of each formal concept without the need to sacrifice the quality of results. GreCon2 for each one in the input matrix maintains a list of formal concepts covering it. Due to these lists, it is no longer necessary to recompute coverage for all formal concepts in step~2. Only coverage of concepts covering some uncovered ones is updated. In essence, this way GreCon2 trades speed for memory. Indeed, it does not need to recompute coverage for each formal concept; however, it has to maintain additional data structures.

\begin{figure}[t]
\centering
$
\mathbf{I} =
\begin{bNiceMatrix}[first-row,first-col]
  & a & b & c & d\\
0 &	1 & 0 & 1 & 1 \\
1 &	0 & 1 & 1 & 0 \\
2 &	0 & 0 & 1 & 1 \\
\end{bNiceMatrix}
$
\caption{Example of a binary matrix.}
\label{fig:ctx00}
\end{figure}

\newcolumntype{C}[1]{>{\centering\arraybackslash}p{#1}}
\newcolumntype{M}[1]{>{\centering\arraybackslash}m{#1}}
\begin{figure}
\linespread{1}
\noindent
\centering
\begin{tikzpicture}

\node[anchor=north west] at (1.5, .4) {$cells$};
\node[anchor=north west] at (0, 0) {
\begin{tabular}{@{}M{1cm}|M{1cm}|M{0cm}}
	\cline{2-2}
	$0,a$ &          & \rule{0pt}{0.5cm} \\ \cline{2-2}
	$0,b$ & $\times$ & \rule{0pt}{0.5cm} \\ \cline{2-2}
	$0,c$ &          & \rule{0pt}{0.5cm} \\ \cline{2-2}
	$0,d$ &          & \rule{0pt}{0.5cm} \\ \cline{2-2}
	$1,a$ & $\times$ & \rule{0pt}{0.5cm} \\ \cline{2-2}
	$1,b$ &          & \rule{0pt}{0.5cm} \\ \cline{2-2}
	$1,c$ &          & \rule{0pt}{0.5cm} \\ \cline{2-2}
	$1,d$ & $\times$ & \rule{0pt}{0.5cm} \\ \cline{2-2}
	$2,a$ & $\times$ & \rule{0pt}{0.5cm} \\ \cline{2-2}
	$2,b$ & $\times$ & \rule{0pt}{0.5cm} \\ \cline{2-2}
	$2,c$ &          & \rule{0pt}{0.5cm} \\ \cline{2-2}
	$2,d$ &          & \rule{0pt}{0.5cm} \\ \cline{2-2}
\end{tabular}
};

\node[anchor=north west] at (3.5, -0.5) {
\begin{tabular}{|M{1cm}|M{0cm}}\cline{1-1}
1 & \rule{0pt}{0.5cm} \\ \cline{1-1}
\end{tabular}
};

\node[anchor=north west] at (3.5, -1.5cm) {
\begin{tabular}{|M{1cm}|M{1cm}|M{1cm}|M{0cm}}\cline{1-3}
0 & 1 & 4 & \rule{0pt}{0.5cm} \\ \cline{1-3}
\end{tabular}
};

\node[anchor=north west] at (3.5, -2.5cm) {
\begin{tabular}{|M{1cm}|M{1cm}|M{0cm}}\cline{1-2}
1 & 4 & \rule{0pt}{0.5cm} \\ \cline{1-2}
\end{tabular}
};

\node[anchor=north west] at (3.5, -3.5) {
\begin{tabular}{|M{1cm}|M{0cm}}\cline{1-1}
3 & \rule{0pt}{0.5cm} \\ \cline{1-1}
\end{tabular}
};

\node[anchor=north west] at (3.5, -4.5cm) {
\begin{tabular}{|M{1cm}|M{1cm}|M{0cm}}\cline{1-2}
0 & 3 & \rule{0pt}{0.5cm} \\ \cline{1-2}
\end{tabular}
};

\node[anchor=north west] at (3.5, -5.5cm) {
\begin{tabular}{|M{1cm}|M{1cm}|M{0cm}}\cline{1-2}
0 & 4 & \rule{0pt}{0.5cm} \\ \cline{1-2}
\end{tabular}
};

\node[anchor=north west] at (3.5, -6.5cm) {
\begin{tabular}{|M{1cm}|M{1cm}|M{0cm}}\cline{1-2}
0 & 4 & \rule{0pt}{0.5cm} \\ \cline{1-2}
\end{tabular}
};

\node[anchor=north west] at (9.85, .4) {$concepts$};
\node[anchor=north west] at (8, 0) {
\begin{tabular}{M{0.5cm}|M{3cm}|M{0cm}}
	\cline{2-2}
	0 & $\langle \{0, 1, 2\}, \{c\}\rangle$  & \rule{0pt}{0.5cm} \\ \cline{2-2}
	1 & $\langle \{0\}, \{a, c, d \}\rangle$ & \rule{0pt}{0.5cm} \\ \cline{2-2}
	2 & $\langle \{\}, \{a,b,c,d \}\rangle$  & \rule{0pt}{0.5cm} \\ \cline{2-2}
	3 & $\langle \{1\}, \{b, c \}\rangle$    & \rule{0pt}{0.5cm} \\ \cline{2-2}
	4 & $\langle \{0, 2\}, \{c, d\}\rangle$  & \rule{0pt}{0.5cm} \\ \cline{2-2}
\end{tabular}
};

\node[anchor=north west] at (9, -3.6) {$covers$};
\node[anchor=north west] at (8, -4) {
\begin{tabular}{M{0.5cm}|M{1cm}|M{0cm}}
	\cline{2-2}
	0 & 3  & \rule{0pt}{0.5cm} \\ \cline{2-2}
	1 & 3  & \rule{0pt}{0.5cm} \\ \cline{2-2}
	2 & 0  & \rule{0pt}{0.5cm} \\ \cline{2-2}
	3 & 2  & \rule{0pt}{0.5cm} \\ \cline{2-2}
	4 & 4  & \rule{0pt}{0.5cm} \\ \cline{2-2}
\end{tabular}
};

\draw[-{Stealth[length=8pt, width=6pt]}] (2.05cm, -.45) -- (3.6,-.99); \filldraw[black] (2.05,-.45) circle (2.5pt);
\draw[-{Stealth[length=8pt, width=6pt]}] (2.05cm, -1.8) -- (3.6,-1.95); \filldraw[black] (2.05,-1.8) circle (2.5pt);
\draw[-{Stealth[length=8pt, width=6pt]}] (2.05cm, -2.45) -- (3.6,-2.95); \filldraw[black] (2.05,-2.45) circle (2.5pt);
\draw[-{Stealth[length=8pt, width=6pt]}] (2.05cm, -3.75) -- (3.6,-3.95); \filldraw[black] (2.05,-3.75) circle (2.5pt);
\draw[-{Stealth[length=8pt, width=6pt]}] (2.05cm, -4.4) -- (3.6,-4.95); \filldraw[black] (2.05,-4.4) circle (2.5pt);
\draw[-{Stealth[length=8pt, width=6pt]}] (2.05cm, -7.0) -- (3.6,-5.95);\filldraw[black] (2.05,-7.0) circle (2.5pt);
\draw[-{Stealth[length=8pt, width=6pt]}] (2.05cm, -7.65) -- (3.6,-6.95);\filldraw[black] (2.05,-7.65) circle (2.5pt);
\end{tikzpicture}
\caption{GreCon2's representation of the binary matrix from Fig.~\ref{fig:ctx00}.}
\label{fig:gc2-structs}
\end{figure}
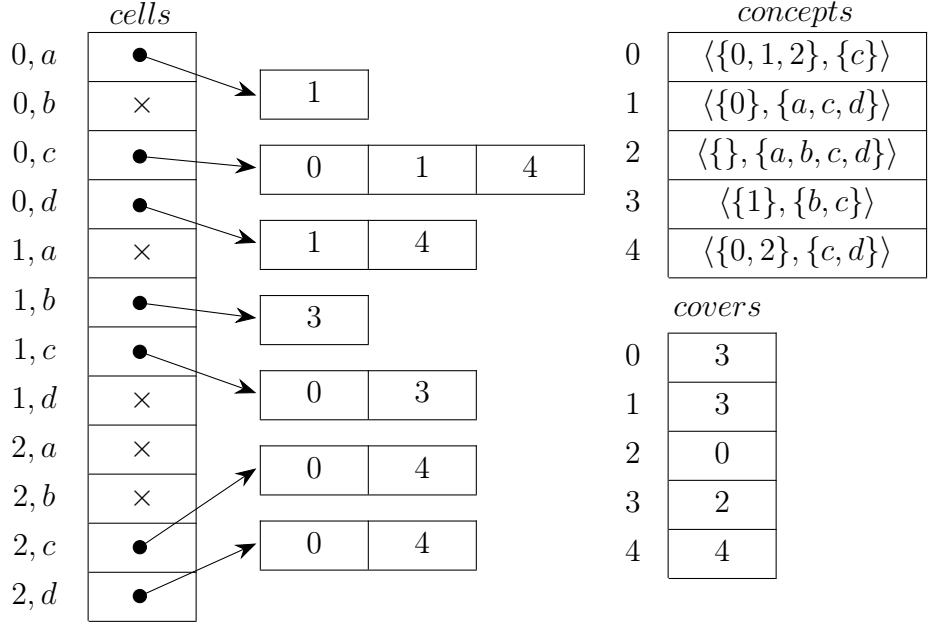

\medskip

We have identified multiple spots where GreCon2 can be improved to obtain a faster algorithm. We are to discuss them and propose improvements. Before we delve into details, we recall GreCon2 and data structures it relies on.

\begin{algorithm}[t]
\caption{\textsc{GreCon2}}
\label{alg:grecon2}
\KwIn{Boolean matrix $\mathbf I \in \{0, 1\}^{m \times n}$.}
\KwOut{Set ${\cal F}$ of factor concepts.}

$concepts \leftarrow  {\cal B}(\mathbf I)$\;\label{alg:gc2:concepts}
${\cal F} \leftarrow \emptyset$\;
\ForEach{$\langle A_l, B_l \rangle \in concept$s}{\label{alg:gc2:initS}
	$covers[l] \leftarrow ||A_l||\cdot ||B_l||$\;
	\ForEach{$i \in A_l$}{
		\ForEach{$j \in B_l$}{
			append $l$ to list $cell[i\cdot n + j]$\;
		}
	}\label{alg:gc2:initE}
}
\While{$\mathbf A_{\cal F} \circ \mathbf B_{\cal F} \not= \mathbf I$}{
$l \leftarrow$ find index $l$ in $covers$ such that $covers[l]$ is maximal\; \label{alg:gc2:max}
$ \langle A_l, B_l \rangle \leftarrow concepts[l]$\;
$ {\cal F} \leftarrow  {\cal F} \cup \{ \langle A_l, B_l \rangle \}$\;  \label{alg:gc2:add}
	\ForEach{$i \in A_l$}{\label{alg:gc2:uncoverS}
		\ForEach{$j \in B_l$}{
			\ForEach{$k \in cell[i \cdot n + j]$}{
				$covers[k] \leftarrow covers[k] -1$\;
			}
			delete list $cell[i \cdot n + j]$
		}
		
	}\label{alg:gc2:uncoverE}
}
\Return $\cal  F$
\end{algorithm}

The GreCon2 algorithm uses three arrays---$concepts, covers$, and $cells$. The $concepts$ array contains all formal concepts; the $covers$ array has the same size as $concepts$  and contains number of ones that each formal concept covers. The $cells$ array corresponds to the input matrix, in that way, that each cell of the array is corresponding to exactly one cell of the input matrix. Each cell in this array contains list of indexes to formal concepts covering the given one. Naturally, $cells$ has size $m \times n$. See Figure~\ref{fig:gc2-structs} for illustration on how a matrix from Fig.~\ref{fig:ctx00} is represented in GreCon2.

At the beginning  GreCon2 (see Algorithm~\ref{alg:grecon2} for pseudo-code) obtains all formal concepts (line~\ref{alg:gc2:concepts}) and initializes all three arrays (lines \ref{alg:gc2:initS}--\ref{alg:gc2:initE}). Then, it finds a formal concept with the largest coverage (line~\ref{alg:gc2:max}), returns it (line~\ref{alg:gc2:add}), and uncovers ones covered by this factor concept (lines \ref{alg:gc2:uncoverS}--\ref{alg:gc2:uncoverE}). These steps are repeated until complete factorization is obtained.

\subsection{Data Structure}
\label{sec:grecon3:struct}

Although using an one dimensional array to represent a matrix is straightforward and might at first sight seem efficient (in terms of accessing individual elements),\footnote{From the theoretical point of view we usually presume that array access is done in a constant time. From the practical viewpoint this is not a true in general. Due to hardware design relying on multiple levels of caches, access times to the main memory of a computer may vary significantly depending on access patterns. Access to the data in a CPU cache may be up to two orders of magnitude faster than access to the data in the main memory.} it is not suitable for sparse datasets which are commonly encountered in practice. In such cases many of the cells are empty which means that a lot of memory is used inefficiently. Further, if values are scattered throughout the memory, it leads to inefficient cache utilization which means slower operations with the matrix. Furthermore, notice that with each iteration the input matrix becomes sparser and sparser.

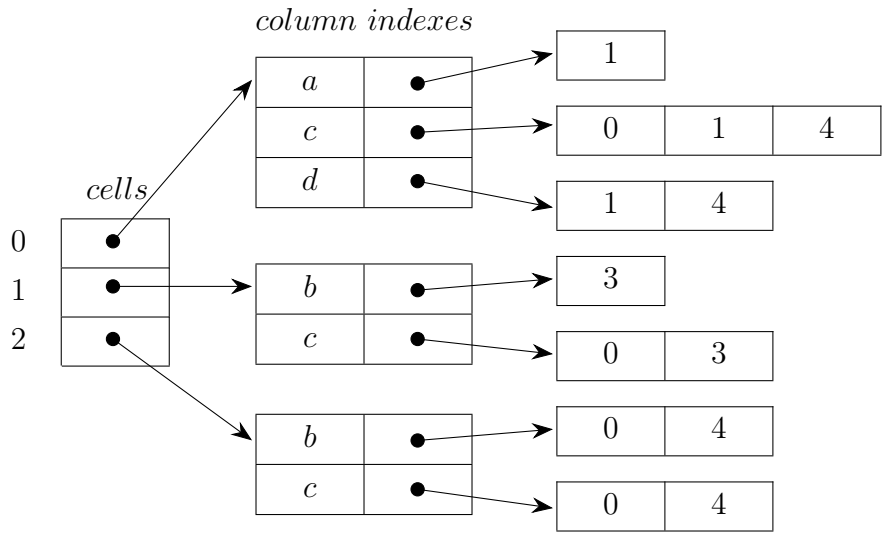
\begin{figure}
\linespread{1}
\noindent
\centering
\begin{tikzpicture}

\node[anchor=north west] at (1.25, .55) {$cells$};
\node[anchor=north west] at (0, 0) {
\begin{tabular}{@{}M{0.7cm}|M{1cm}|M{0cm}}
	\cline{2-2}
	$0$ &          & \rule{0pt}{0.5cm} \\ \cline{2-2}
	$1$ &  & \rule{0pt}{0.5cm} \\ \cline{2-2}
	$2$ &          & \rule{0pt}{0.5cm} \\ \cline{2-2}
\end{tabular}
};

\node[anchor=north west] at (3.5, 2.8) {
$column$ $indexes$
};
\node[anchor=north west] at (3.5, 2.15) {
\begin{tabular}{|M{1cm}|M{1cm}|} \hline
$a$ &  \rule{0pt}{0.5cm} \\ \hline
$c$ &  \rule{0pt}{0.5cm} \\ \hline
$d$ &  \rule{0pt}{0.5cm} \\ \hline
\end{tabular}
};

\node[anchor=north west] at (3.5, -0.6) {
\begin{tabular}{|M{1cm}|M{1cm}|} \hline
$b$ &  \rule{0pt}{0.5cm} \\ \hline
$c$ &  \rule{0pt}{0.5cm} \\ \hline
\end{tabular}
};

\node[anchor=north west] at (3.5, -2.6) {
\begin{tabular}{|M{1cm}|M{1cm}|} \hline
$b$ &  \rule{0pt}{0.5cm} \\ \hline
$c$ &  \rule{0pt}{0.5cm} \\ \hline
\end{tabular}
};

\node[anchor=north west] at (7.5, 2.5) {
\begin{tabular}{|M{1cm}|M{0cm}}\cline{1-1}
1 & \rule{0pt}{0.5cm} \\ \cline{1-1}
\end{tabular}
};

\node[anchor=north west] at (7.5, 1.5cm) {
\begin{tabular}{|M{1cm}|M{1cm}|M{1cm}|M{0cm}}\cline{1-3}
0 & 1 & 4 & \rule{0pt}{0.5cm} \\ \cline{1-3}
\end{tabular}
};

\node[anchor=north west] at (7.5, 0.5cm) {
\begin{tabular}{|M{1cm}|M{1cm}|M{0cm}}\cline{1-2}
1 & 4 & \rule{0pt}{0.5cm} \\ \cline{1-2}
\end{tabular}
};

\node[anchor=north west] at (7.5, -0.5) {
\begin{tabular}{|M{1cm}|M{0cm}}\cline{1-1}
3 & \rule{0pt}{0.5cm} \\ \cline{1-1}
\end{tabular}
};

\node[anchor=north west] at (7.5, -1.5cm) {
\begin{tabular}{|M{1cm}|M{1cm}|M{0cm}}\cline{1-2}
0 & 3 & \rule{0pt}{0.5cm} \\ \cline{1-2}
\end{tabular}
};

\node[anchor=north west] at (7.5, -2.5cm) {
\begin{tabular}{|M{1cm}|M{1cm}|M{0cm}}\cline{1-2}
0 & 4 & \rule{0pt}{0.5cm} \\ \cline{1-2}
\end{tabular}
};

\node[anchor=north west] at (7.5, -3.5cm) {
\begin{tabular}{|M{1cm}|M{1cm}|M{0cm}}\cline{1-2}
0 & 4 & \rule{0pt}{0.5cm} \\ \cline{1-2}
\end{tabular}
};

\draw[-{Stealth[length=8pt, width=6pt]}] (1.75cm, -.45) -- (3.6,1.7); \filldraw[black] (1.75,-.45) circle (2.5pt);

\draw[-{Stealth[length=8pt, width=6pt]}] (1.75cm, -1.05) -- (3.6,-1.05); \filldraw[black] (1.75,-1.05) circle (2.5pt);

\draw[-{Stealth[length=8pt, width=6pt]}] (1.75cm, -1.75) -- (3.6,-3.1); \filldraw[black] (1.75,-1.75) circle (2.5pt);


\draw[-{Stealth[length=8pt, width=6pt]}] (5.8, 1.65) -- (7.6,2.05); \filldraw[black] (5.8,1.65) circle (2.5pt);
\draw[-{Stealth[length=8pt, width=6pt]}] (5.8, 1.0) -- (7.6,1.1); \filldraw[black] (5.8,1.0) circle (2.5pt);
\draw[-{Stealth[length=8pt, width=6pt]}] (5.8, 0.35) -- (7.6,0.0); \filldraw[black] (5.8,0.35) circle (2.5pt);

\draw[-{Stealth[length=8pt, width=6pt]}] (5.8, -1.1) -- (7.6,-0.95); \filldraw[black] (5.8,-1.1) circle (2.5pt);
\draw[-{Stealth[length=8pt, width=6pt]}] (5.8, -1.75) -- (7.6,-1.95); \filldraw[black] (5.8,-1.75) circle (2.5pt);
\draw[-{Stealth[length=8pt, width=6pt]}] (5.8, -3.1) -- (7.6,-2.95); \filldraw[black] (5.8,-3.1) circle (2.5pt);
\draw[-{Stealth[length=8pt, width=6pt]}] (5.8, -3.75) -- (7.6,-4.00); \filldraw[black] (5.8,-3.75) circle (2.5pt);

\end{tikzpicture}
\caption{Example of a sparse representation of the $cells$ array for input matrix from Fig.~\ref{fig:ctx00}.}
\label{fig:gc3-structs}
\end{figure}

To resolve this weakness we propose to use data structure inspired by sparse matrix representation. In fact, it can be seen as a two dimensional jagged array. Each row of the input matrix is represented with a cell in the $cells$ array. Each cell contains an array\footnote{Strictly speaking it is a reference to an array.} which is a sequence of pairs $\langle \mathit{column}, \mathit{indexes}\rangle$. See Figure~\ref{fig:gc3-structs}, for an illustration. This data structure allows to reduce memory footprint of the algorithm and improve cache utilization. Benefits of this representation are most evident with sparse data. In fact, further optimizations discussed in the present paper aim to keep the $cells$  array as sparse as possible.

\begin{remark}
Even though the proposed data structure is different than the $cells$ array in GreCon2, its purpose is the same, i.e.\ it keeps a track on which concepts covers particular one in the input matrix. Thus, in what follows we keep call it `the $cells	$ array'.
\end{remark}

\subsection{Initialization}
\label{sec:grecon3:init}

The GreCon2 algorithm spends a substantial amount of time in the initialization phase of the algorithm (see Algorithm~\ref{alg:grecon2}, lines \ref{alg:gc2:initS}--\ref{alg:gc2:initE}). Basically, for each concept and each one it covers, it has to store an index into a corresponding list. If large numbers of formal concepts are involved or if they cover large quantities of ones, this may constitute a performance bottleneck. Moreover, it shows up that some indexes may not be relevant for  a given computation. For example, for the matrix in Figure~\ref{fig:ctx01}, when the first two factors $c_1$ and  $c_2$ are discovered, it is clearly evident that formal concept $c_4$ is not a factor concept and can be ignored. However, GreCon2 spends CPU time and memory to setup and maintain information on this concept anyways.

We propose not to initialize the $cells$, $covers$, and $concepts$ arrays \textit{en bloc}, rather fill in values \textit{on the fly} as relevant concepts appears. Further, it makes sense to focus on large concepts first, i.e.\ concepts having greatest size $||A||\cdot||B||$. Naturally, large concept tends to cover more ones, hence provides better explanation for the input data. Therefore, we are to assume that formal concepts are totally ordered by their size, largest concepts first.\footnote{To resolve ties, one may prefer concepts having larger intent or extent, use lexicographical ordering of attributes, etc.} This ensures that larger concepts are processed first.

Under this assumption, we take an ordered list of formal concepts and one by one consider them as possible factor concepts.  For each formal concept its actual coverage is computed and corresponding lists of indexes in the $cells$ array are updated. It stops, whenever the formal concept has a smaller size than the formal concept having the best coverage so far. Obviously, there are no other formal concepts that could cover more ones than those already encountered. Subsequently, a factor concept (with the highest coverage) is returned. Ones covered by this concept are turned to zeros, and coverage of relevant concepts is adjusted the same way GreCon2 does.


This enhancement means two changes to the algorithm: (i) We have to change the way concepts are retrieved. This is discussed in Section~\ref{sec:grecon3} where the algorithm with all optimizations is presented. (ii) Initialization of the $cells$ array has to be adjusted.

To accommodate this change we have separated initialization of the $cells$ array into a procedure \textsc{CoverConcept}, see Algorithm~\ref{alg:cover-cencept} for its pseudo-code. This procedure takes a formal concept an fills into lists in $cells$ index of the given concept. In fact, this procedure is similar to the initialization code of GreCon2 (cf. Algorithm~\ref{alg:grecon2}, lines~\ref{alg:gc2:initS}--\ref{alg:gc2:initE}). The \textsc{ConceptCover} procedure, checks if the $cells$ array has a reference to a valid list at the given coordinates (line~\ref{alg:concept-cover:check}, $n$ is the number of columns in the input matrix $\mathbf I$). If not, the cell is ignored since it was covered by some other factor concept. Further, it has to count how many cells is covered by this concepts (lines~\ref{alg:concept-cover:cov1} and~\ref{alg:concept-cover:cov2}).

\begin{algorithm}
\caption{\textsc{CoverConcept}($\langle A, B \rangle, l$)}
\KwIn{Formal concept $\langle A, B \rangle$ that is considered as a candidate concept and its index in the $concepts$ array.}
\KwOut{Number of ones the given formal covers.}
\label{alg:cover-cencept}
$cover \leftarrow 0$\;
\ForEach{$i \in A$}{
	\ForEach{$j \in B$}{
		\If{$cell[i\cdot n + j]$ {\bf is not empty}} {\label{alg:concept-cover:check}
			append $l$ to list $cell[i\cdot n + j]$\;
			$cover \leftarrow cover + 1$\;\label{alg:concept-cover:cov1}
		}
	}
}
$covers[l] \leftarrow cover$\;\label{alg:concept-cover:cov2}
\Return $cover$\;
\end{algorithm}

This optimization, which postpones initialization of the $cells$ array, has multiple notable consequences:
\begin{itemize}
\item Only relevant concepts are kept in the memory. This reduces memory requirements of the algorithm.
\item Time spent in the initialization phase is reduced. This has also a positive impact if approximate factorization is expected, because small and unused concepts are simply ignored.
\item Only ones that have not been covered so far, have their lists updated. This substantially reduces number of CPU operations to maintain these lists, if compared with GreCon2.
\end{itemize}

\begin{figure}[h!]
\centering
$
\mathbf{I} =
\begin{bNiceMatrix}[first-row,first-col]
      & a & b & c & d & e & f & g \\
0 & 1 & 1 & 1 & 1 & 0 & 0 & 0 \\
1 & 1 & 1 & 1 & 1 & 0 & 0 & 0 \\
2 & 1 & 1 & 1 & 1 & 1 & 1 & 1 \\
3 & 0 & 0 & 1 & 1 & 1 & 1 & 1 \\
4 & 0 & 0 & 1 & 1 & 0 & 0 & 0
\end{bNiceMatrix}
$
\hspace{20pt}
$
\begin{array}{ll}
c_1 = & \langle \{ 0, 1, 2 \} ; \{a, b, c, d\} \rangle \\
c_2 = & \langle \{ 2, 3 \} ; \{c, d, e, f, g\} \rangle \\
c_3 = & \langle \{ 0, 1, 2, 3, 4 \} ; \{c, d\} \rangle \\
c_4 = & \langle \{ 2 \} ; \{a, b, c, d, e, f, g\} \rangle \\
\\
\end{array}
$
\caption{Example of the binary matrix (left) and formal concepts in this matrix (right).}
\label{fig:ctx01}
\end{figure}

\begin{figure}[h!]
\centering
\[
\mathbf{I} - (\mathbf{A}_{\mathcal{F}} \circ \mathbf{B}_{\mathcal{F}}) =
\begin{bNiceMatrix}[first-row,first-col]
      & a & b & c & d & e & f & g \\
0 & 0 & 0 & 0 & 0 & 0 & 0 & 0 \\
1 & 0 & 0 & 0 & 0 & 0 & 0 & 0 \\
2 & 0 & 0 & \mathbf{0} & \mathbf{0} & \mathbf{1} & \mathbf{1} & \mathbf{1} \\
3 & 0 & 0 & \mathbf{1} & \mathbf{1} & \mathbf{1} & \mathbf{1} & \mathbf{1} \\
4 & 0 & 0 & 1 & 1 & 0 & 0 & 0
\end{bNiceMatrix}
\;=\;
\begin{bNiceMatrix}[first-row,first-col]
      & a & b & c & d & e & f & g \\
0 & 0 & 0 & \mathbf{0} & \mathbf{0} & 0 & 0 & 0 \\
1 & 0 & 0 & \mathbf{0} & \mathbf{0} & 0 & 0 & 0 \\
2 & 0 & 0 & \mathbf{0} & \mathbf{0} & 1 & 1 & 1 \\
3 & 0 & 0 & \mathbf{1} & \mathbf{1} & 1 & 1 & 1 \\
4 & 0 & 0 & \mathbf{1} & \mathbf{1} & 0 & 0 & 0
\end{bNiceMatrix}
\]

\caption{Binary matrix from Fig.~\ref{fig:ctx01} after discovery of the factor concept $c_1$ (i.e., $\mathcal F = \{c1\}$) and highlighted concepts $c_2$ (left) and $c_3$ (right).}
\label{fig:ctx02}
\end{figure}

%

\subsection{Incremental Coverage Computation}

The point where the arrays $cells$ and $covers$ are fully initialized with some concept may be pushed even further. For instance, consider the input matrix from Fig.~\ref{fig:ctx01} and assume that the first factor $c_1$ was discovered, see Fig.~\ref{fig:ctx02}. Notice that concepts $c_2$ and $c_3$ have the same size (i.e.\ $10$). If  $cells$ and $covers$ are initialized with formal concept $c_2$, we obtain information that it covers $8$ ones not covered yet. Afterwards, we perform initialization with $c_3$ and obtain that it covers $4$ ones not covered yet. Since formal concept $c_4$ has a smaller size than $c_2$ and $c_3$, initialization phase of the algorithm is stopped, and formal concept $c_2$ is selected as the second factor concept.

Let us take a closer look at the initialization of the formal concept $c_3$. If we iterate over $\mathbf I_{ij}$ by rows (rows from  the top to the bottom and each row from the left to the right), i.e., we check $\mathbf I_{0c}, \mathbf I_{0d}, \mathbf I_{1c}, \mathbf I_{1d}, \ldots, \mathbf I_{4d}$. We do not have to initialize $cells$ for $\mathbf I_{0c}, \mathbf I_{0d}, \ldots, \mathbf I_{2d}$, since these cells are already covered by the factor concept $c_1$. Moreover, after processing $\mathbf I_{1c}$ it is clear that coverage of $c_3$ is less than $8$ (since $\mathbf I_{0c}=\mathbf I_{0d}=\cdots=\mathbf I_{2d} = 0$). This means $c_3$ is definitely a worse factor concept than $c_2$, and thus, the initialization of $c_3$ can be suspended at this point of algorithm's execution. Notice that when uncovering $c_2$ we do not have to manipulate with cells shared with the formal concept $c_3$ (i.e., with cells $\mathbf I_{3c}$ and $\mathbf I_{3d}$) because they were not initialized with indexes of $c_3$.

Our intention is to initialize $cells$ and $covers$ incrementally. If it is clear, that a formal concept can not cover more ones than some formal concept encountered earlier, initialization is suspended. Later, if a formal concept seems to be a relevant candidate, its initialization is resumed.

To achieve this, we modified the $\textsc{CoverConcept}$ procedure to cover ones row by row. If it is apparent that the coverage of the currently processed concept is worse than the coverage of some previously processed one, it stops. To enable this functionality, we introduce two new arrays---$\mathit{potential}$ and $\mathit{progress}$. Their purpose is to keep a track on number of ones that formal concept may potentially cover (if unprocessed rows are also considered), and the last row that was processed during initialization, respectively.

In other words, value $covers[l] + potential[l]$ represents an upper bound of ones that formal concept with index $l$ may cover. Initially, $potential[l] = ||A_l|| \cdot||B_l||$, then its value is decreasing to $potential[l] = 0$. In that case $covers[l]$ is the exact number of ones covered by a formal concept $l$. Value $progress[l]$ is the point (row) where computation of a coverage has been stopped and where it should be resumed. See pseudo-code in Algorithm~\ref{alg:cover-incremental}. This procedure is similar to \textsc{CoverConcept} (see Algorithm~\ref{alg:cover-cencept}) with several changes.

\begin{itemize}
\item It gets the last known coverage of the input formal concept (line~\ref{alg:cover-incremental:getcov}).
\item It ignores  already processed rows (line~\ref{alg:cover-incremental:cond}). Recall that this information is kept in the $progress$ array.
\item When each row is processed, potential of the concept to cover ones is adjusted (line~\ref{alg:cover-incremental:pot}).
\item It checks if a concept is a relevant candidate, i.e. its obtained coverage and its potential to cover ones is not less than the best coverage known so far (line~\ref{alg:cover-incremental:check}). If the formal concept is still relevant, the procedure proceeds with processing of the next row. Otherwise, the last processed row is stored in the $progress$ array (line \ref{alg:cover-incremental:store}) and the procedure stops (line~\ref{alg:cover-incremental:break}).
\end{itemize}

\begin{remark}
The potential of formal concept to cover ones could be checked in the inner loop of the algorithm. This would provide more fine-grained control of coverage computation, however, it would also mean a performance penalty. Therefore, we perform check after processing each row (line~\ref{alg:cover-incremental:pot}).
\end{remark}

\begin{algorithm}[ht]
\small
\caption{\textsc{CoverIncremental}($\langle A, B \rangle, l, bestCoverage$)}
\label{alg:cover-incremental}
\KwIn{Formal concept $\langle A, B \rangle$ that is considered as a candidate concept and its index $l$ in the $concepts$ array. Minimal number of ones (parameter $bestCoverage$) concept has to cover to be considered as relevant factor concept.}
\KwOut{Number of ones the given formal covers.}
$cover \leftarrow covers[l]$\;\label{alg:cover-incremental:getcov}
\ForEach{$i \in A$ {\bf such that} $i > progress[l]$}{\label{alg:cover-incremental:cond}
	\ForEach{$j \in B$}{
		\If{$cell[i\cdot n + j]$ {\bf is not empty}} {
			append $l$ to list $cell[i\cdot n + j]$\;
			$cover \leftarrow cover + 1$\;
		}
	}
	$potential[l] \leftarrow  potential[l] - ||B||$\;\label{alg:cover-incremental:pot}
	\If{$covers[l] + potential[l] < bestCoverage$}{\label{alg:cover-incremental:check}
		$progress[l] \leftarrow i$\;\label{alg:cover-incremental:store}
		{\bf break};\label{alg:cover-incremental:break}
	}
}
$covers[l] \leftarrow cover$\;
\Return $cover$\;
\end{algorithm}

\subsection{Prominent Cases}

The GreCon2 algorithm treats all factors equally. Since we assume that all formal concepts are ordered from the largest one to the smallest one, we may get some factors more efficiently.

\subsubsection{First Factor Discovery}
\label{sec:grecon3:firstconcept}
Assuming that formal concepts on input are ordered w.r.t.\ their sizes, it is truly easy to identify the first factor concept. It is the first formal concept on input. This trivial observation has two practical implications. We can obtain the first factor concept directly without the need to compute its coverage from the input matrix, or other steps that GreCon2 does. Furthermore, since the first factor is the largest one, it usually tends to cover high amount of ones that are immediately uncovered in the first step of the algorithm. Hence, it reduces number of values to setup in the $cells$ array. This, along with the aforementioned optimizations (Sec.~\ref{sec:grecon3:struct} and~\ref{sec:grecon3:init}), has also a positive performance impact on discovery of further factors.

\subsubsection{Second Factor Discovery}
\label{sec:grecon3:secondconcept}
When looking for the second factor concept, we can also avoid computing its coverage from the input matrix. In this case we can not solely rely on the order of formal concepts, since we have to take into account that some ones are already covered by the first factor concept. On the other hand, we may use the fact that the coverage of the second factor concept is given as its size minus number of ones that are common with the first factor, i.e., its number of  objects times number of attributes the given formal concept shares with the first factor concept. Formally, let $\langle A, B\rangle$ be the first formal concept, then the coverage of a formal concept $\langle C, D\rangle$ is given as
$$cov(\langle C, D\rangle) = ||C|| \cdot ||D|| - ||A \cap C||\cdot ||B \cap D||.$$
Basically, instead of checking $||C|| \cdot ||D||$ values in the input matrix, we just compute two set intersections. Since the second factors also tend to be large, this optimization has very similar positive performance implications as the optimization for the first factor.

\subsubsection{Third and Further Factor Discovery}
\label{sec:grecon3:thirdconcept}
We may use similar approach to find the third factor concept. Let $\langle A, B\rangle$ and $\langle C, D\rangle$ be the first two factor concepts then the coverage of the formal concept $\langle E, F\rangle$ is given as 
\begin{align*}
\operatorname{cov}(\langle E, F\rangle)
&= ||E|| \cdot ||F||
   - ||A \cap E|| \cdot ||B \cap F|| -\\
&\quad
   - ||C \cap E|| \cdot ||D \cap F||
   + ||A \cap C \cap E|| \cdot ||B \cap D \cap F||.
\end{align*}

Using the inclusion-exclusion principle it could be possible to get coverage for the fourth, fifth, and further concepts. From our experience such optimization does not bring significant performance improvement while complicating algorithm implementation. Therefore, we use optimization only for the first three concepts and then we use the default strategy proposed in GreCon2. 

\subsection{GreCon3}
\label{sec:grecon3}

We have incorporated discussed optimizations into GreCon2. This resulted in a novel algorithm we call GreCon3. Its pseudo-code is presented as Algorithm~\ref{alg:grecon3},  \ref{alg:loadconcepts}, \ref{alg:cov}, \ref{alg:cover-cencept}, \ref{alg:cover-incremental}, and \ref{alg:uncov}. For better comprehension the algorithm is split into six procedures each performing a specific task.

\subsubsection{GreCon3 main procedure}

The pseudo-code of the main procedure of the GreCon3 algorithm is presented as Algorithm~\ref{alg:grecon3}. This procedure accepts matrix~$\mathbf I$ to decompose and a sorted list of formal concepts $\mathcal B^*$ obtained by any suitable algorithm for computing formal concepts, for instance, from the CbO family~\cite{KoKr:50shadesOfCbO}. 

First, all auxiliary data structures are initialized and empty (line~\ref{alg:gc3:init}). For brevity, we assume their size may be extended during the algorithm execution and all of them are in a global scope, i.e., any procedure can access them.

Meaning of $concepts$, $cells$, $covers$, $potential$ is explained in the previous sections. Beside these GreCon3 uses an auxiliary queue $\mathcal Q$ containing indexes of concepts that were already read from the list $\mathcal B^*$. In essence, formal concepts in the queue $\mathcal Q$ are the largest concepts obtained, and thus, are serving as candidates for factor concepts.

Immediately afterwards, first concept is discovered (lines~\ref{alg:gc3:fc1S} and \ref{alg:gc3:fc1E}). Core of the algorithm lies in the loop (lines \ref{alg:gc3:loopS}--\ref{alg:gc3:loopE}) checking if the factorization has been obtained (line~\ref{alg:gc3:loopS}) and if not, auxiliary procedure \textsc{LoadConcepts} is called (line~\ref{alg:gc3:load}). This procedure has multiple intertwined roles. It loads concepts from the input list $\mathcal B^*$, place them into $\mathcal Q$, computes their coverage, and selects a factor concept with the best coverage whose index it returns. We explain this procedure in detail in what follows.

Afterwards, ones that are covered by this concept (line~\ref{alg:gc3:select}) are uncovered (line~\ref{alg:gc3:uncover}) and factor concept is placed into the set $\mathcal F$.

Further, values in the queue $\mathcal Q$ are reordered so that concepts whose sum $covers[l] + potential[l]$ is higher are placed first (line~\ref{alg:gc3:sort}). Thanks to this the \textsc{LoadConcepts} procedure tries concepts covering more ones (or at least having a potential to cover more ones) first. Then, concepts that are no longer covering any one are removed (line~\ref{alg:gc3:remove}). Since the queue $\mathcal Q$ was ordered in the preceding step, such concepts can be found at the end of the queue.

Loop stops when all ones are covered and factor concepts are returned (line~\ref{alg:gc3:ret}).

Note that when the third factor is discovered, we cannot rely on optimizations for the first three factors anymore. Hence, the $cells$ arrays has to be initialized (lines \ref{alg:gc3:initCellsS}--\ref{alg:gc3:initCellsE}). Basically, into each cell that corresponds to an uncovered one is insert an empty list. These lists are subsequently filled in the \textsc{LoadConcepts} procedure.

\begin{algorithm}[ht]
\small
\caption{\textsc{GreCon3}($\mathbf I, \mathcal B^*$)}
\label{alg:grecon3}
\KwIn{Boolean matrix $\mathbf I \in \{0, 1\}^{m \times n}$. Sorted list of formal concepts $\mathcal B^*$.}
\KwOut{Set $\cal F$ of factor concepts.}

initialize empty arrays $concepts$, $cells$, $covers$, $potential$, and $\mathcal Q$\;\label{alg:gc3:init}
$\langle A_0, B_0\rangle \leftarrow$ read the first concept from $\mathcal B^*$\;\label{alg:gc3:fc1S}
${\cal  F} \leftarrow \{\langle A_0, B_0\rangle\}$\;\label{alg:gc3:fc1E}

\While{$\mathbf A_{\cal  F} \circ \mathbf B_{\cal  F} \not= \mathbf I$}{\label{alg:gc3:loopS}
	\If{$||\mathcal F|| = 3$}{\label{alg:gc3:initCellsS}
		\ForEach{$i,j$ \rm such that $\mathbf (I - \mathbf A_{\cal  F} \circ \mathbf B_{\cal  F})_{ij} = 1$}{
			$cells[i \cdot n + j] \leftarrow$ empty list\;\label{alg:gc3:initCellsE}
		}
	}
	$l\leftarrow $ \textsc{LoadConcepts}$(\mathcal B^*, \mathcal F)$\;\label{alg:gc3:load}
	$\langle A_l, B_l \rangle \leftarrow concepts[l]$\;\label{alg:gc3:select}
	\textsc{Uncover}($\langle A_l, B_l \rangle$)\;\label{alg:gc3:uncover}
	$ {\cal F} \leftarrow  {\cal F} \cup \{ \langle A_l, B_l \rangle \}$\;\label{alg:gc3:result}
	sort values in $\mathcal Q$ by $covers[l] + potential[l]$\;\label{alg:gc3:sort}
	remove from $\mathcal Q$ values with $covers[l] + potential[l] = 0$\;\label{alg:gc3:loopE}\label{alg:gc3:remove}
}
\Return{$\cal F$}\;\label{alg:gc3:ret}
\end{algorithm}

\begin{remark}
The algorithm may be modified to provide approximate (from-below) decomposition. It is sufficient to replace the loop condition (line~\ref{alg:gc3:loopS}), for instance, with $\frac{||\mathbf A_{\cal  F} \circ \mathbf B_{\cal  F}||}{||\mathbf I||} < \varepsilon$, where $\varepsilon$ is a parameter from the real unit interval $(0; 1]$ indicating the ratio of ones that has to be covered to the total number of ones in the input matrix.
\end{remark}

\subsubsection{Reading concepts}

The GreCon3 algorithm aims to postpone initialization of data structures as much as possible. This means that concept reading and finding the best factor concept must be performed together in a single procedure. Furthermore, since we compute coverage incrementally, we have two lists of concepts (queue $\mathcal Q$ and list of input concepts $\mathcal B^*$) where to look for factor concept candidates. These lists have to be treated somewhat similarly though in specific aspects differently.  See the pseudo-code of the \textsc{LoadConcepts} procedure in Algorithm~\ref{alg:loadconcepts} for details.

The \textsc{LoadConcepts} procedure maintains two variables $bestCoverage$ and $bestConcept$ (lines~\ref{alg:lc:best1} and~\ref{alg:lc:best2}) containing the highest number of ones covered by some concept that was processed so far and index of such concept, respectively.

First, it iterates over concept indexes in the queue $\mathcal Q$ (lines \ref{alg:lc:loop1S}--\ref{alg:lc:loop1E}). For each concept its coverage is obtained using the \textsc{Cover} procedure (line~\ref{alg:lc:cover1}). If its coverage is higher than the coverage of all concepts in the queue processed so far (line~\ref{alg:lc:check1}), variables $bestCoverage$ and $bestConcept$ are adjusted accordingly (lines~\ref{alg:lc:best1adj} and~\ref{alg:lc:best2adj}).

Recall that concepts in this queue are sorted w.r.t.\ to sum $covers[l] + potential[l]$ (see Algorithm~\ref{alg:grecon3}, line~\ref{alg:gc3:sort}). This means if a formal concept with $covers[l] + potential[l]$ lesser than $bestCoverage$ is encountered, it means there are no more concepts in $\mathcal Q$ having better coverage. Hence, the loop is interrupted (line~\ref{alg:lc:break1}).

Subsequently, the \textsc{LoadConcepts} procedure checks the list $\mathcal B^*$ for candidate concepts (lines \ref{alg:lc:loop2S}--\ref{alg:lc:loop2E}). Since we are dealing with new concepts, we have to allocate space in the $concepts$ array and setup its values in $covers$, $progress$, and $potential$ arrays (lines~\ref{alg:lc:initConS}--\ref{alg:lc:initConE}). Since it is a new concept, we do not have any information on its coverage and its potential to cover ones, Therefore these values are set to zero and $||A||\cdot||B||$, respectively. Further, computation of its coverage has not been started yet, thus value of $progress[l]$ is set to be lower than the least column. When the initialization is complete, concept is placed into the queue $\mathcal Q$.

Then, coverage of the concept is obtained (line~\ref{alg:lc:cover2}) and checked against the best coverage obtained so far (lines \ref{alg:lc:check2S}--\ref{alg:lc:check2E}).

If the size of a concept is smaller then $bestCoverage$ we exit the loop (line~\ref{alg:lc:break2}), because concepts in $\mathcal B^*$ are ordered w.r.t.\ their sizes and further unprocessed concepts in $\mathcal B^*$ can not have better coverage.

The procedure returns index in the $concepts$ array of the concept with the highest coverage (line~\ref{alg:lc:ret}).

\begin{remark}
The GreCon3 algorithm does not have to keep all formal concepts in the memory (unlike GreCon and GreCon2). Whenever a concept coverage and its potential to cover ones reaches zero, its slot (index) is freed and available for further use. Therefore, slots in $concepts$, $covers$, and $potential$ arrays can be reused by different concepts. This allows to save memory. In practical implementation it may be appropriate to maintain a list of slots that were freed to obtain index of a free slot efficiently.
\end{remark}

\begin{algorithm}[ht]
\small
\caption{\textsc{LoadConcepts}($\mathcal B^*, \mathcal F$)}
\label{alg:loadconcepts}
\KwIn{Sorted list of formal concepts $\mathcal B^*$. Discovered set of factors~$\mathcal F$.}
\KwOut{Index of the concept with the highest coverage.}

$bestCoverage \leftarrow -1$\;\label{alg:lc:best1}
$bestConcept \leftarrow empty$\;\label{alg:lc:best2}

\While{$\mathcal Q$ contains concept indexes}{\label{alg:lc:loop1S}
	$l \leftarrow$ read next value from $\mathcal Q$\;
	$c \leftarrow $ \textsc{Cover}($concepts[l]$, $\mathcal F$, $bestCoverage$)\;\label{alg:lc:cover1}
	\If{$c > bestCoverage$}{\label{alg:lc:check1}
		$bestConcept \leftarrow l$\;\label{alg:lc:best1adj}
		$bestCoverage \leftarrow c$\;\label{alg:lc:best2adj}
	}
	\lIf{$covers[l] + potential[l] < bestCoverage$}{{\bf break loop}}\label{alg:lc:break1}\label{alg:lc:loop1E}
}

\While{\rm($\mathcal B^*$ contains concepts)}{\label{alg:lc:loop2S}
	$\langle A, B\rangle \leftarrow$ read next concept from $\mathcal B^*$\;
	$l \leftarrow$ empty slot in $concepts$\;
	$covers[l] \leftarrow 0$\;\label{alg:lc:initConS}
	$potential[l] \leftarrow ||A||\cdot||B||$\;
	$concepts[l] \leftarrow \langle A, B\rangle$\;
	$progress[l] \leftarrow -1$\;
	append $l$ to queue $\mathcal Q$\;\label{alg:lc:initConE}
	\lIf{$||A||\cdot||B|| < bestCoverage$}{{\bf break loop}}\label{alg:lc:break2}
	
	$c \leftarrow $ \textsc{Cover}($concepts[l]$, $\mathcal F, bestCoverage$)\;\label{alg:lc:cover2}
	
	\If{$c > bestCoverage$}{\label{alg:lc:check2S}
		$bestConcept \leftarrow l$\;
		$bestCoverage \leftarrow c$\;\label{alg:lc:check2E}
	}
	\label{alg:lc:loop2E}
}
\Return $bestConcept$\;\label{alg:lc:ret}
\end{algorithm}

\subsubsection{Computing coverage}

In the description of the \textsc{LoadConcepts} procedure we mentioned the \textsc{Cover} procedure which takes care of computing formal concept's coverage. Its pseudo-code is shown as Algorithm~\ref{alg:cov}. How the coverage is obtained, depends on the number of factor concepts discovered so far. If only a single factor has been discovered, optimization from Section~\ref{sec:grecon3:secondconcept} is applied and coverage is computed directly (lines \ref{alg:cover:f2S}--\ref{alg:cover:f2E}). If two factors were discovered, coverage is obtained based on optimization from Section~\ref{sec:grecon3:thirdconcept} (lines \ref{alg:cover:f3S}--\ref{alg:cover:f3E}).

In the remaining cases strategy of the GreCon2 algorithm is used (lines \ref{alg:cover:fOS}--\ref{alg:cover:fOE}). This means \textsc{Cover} procedure initializes values in the $cells$ and $covers$ arrays. Subsequently, coverage is obtained from the array $covers$ (line~\ref{alg:cover:fOE}).

Notice that two different procedures are called to initialize $cells$ and $covers$ arrays (see Algorithm~\ref{alg:cover-cencept} and~\ref{alg:cover-incremental} for their pseudo-codes). Based on experience with the algorithm, it turns out that it is not beneficial to use incremental coverage computation for all concepts. We have to take into account that incremental computation has its overhead (memory, slots in $\mathcal Q$, repeated calls of the \textsc{Cover} procedure, etc.). Therefore, if a concept is small (e.g., has less 100 objects) it is initialized \textit{en bloc}, the same way GreCon2 does. Otherwise, incremental initialization is used. In particular this enhancement allows to eliminate concepts that does not cover any one.

\begin{algorithm}[ht]
\small
\caption{\textsc{Cover}($\langle A_l, B_l \rangle,  {\cal F}, bestCoverage$)}
\KwIn{Formal concept $\langle A_l, B_l\rangle$ with unknown coverage. Set of factor concepts $\cal  F$ discovered so far. Highest coverage $bestCoverage$ obtained so far.}
\KwOut{Covarage of the given concept.}
\label{alg:cov}
\Switch{$||\cal  F||$}{
\Case{1}{\label{alg:cover:f2S}
	$\langle A_0, B_0 \rangle \leftarrow$ concept from $\cal  F$\;
	\Return $||A_l||\cdot||B_l|| - ||A_0 \cap A_l||\cdot||B_0 \cap B_l||$\;\label{alg:cover:f2E}
}
\Case{2}{\label{alg:cover:f3S}
	$\langle A_0, B_0 \rangle, \langle A_1, B_1 \rangle \leftarrow$ concepts from $\cal  F$\;
	\Return $||A_l|| \cdot ||B_l|| - ||A_0 \cap A_l||\cdot ||B_0 \cap B_l|| - ||A_1 \cap A_l||\cdot ||B_1 \cap B_l|| + ||A_0 \cap A_1 \cap A_l||\cdot ||B_0 \cap B_1 \cap B_l||$\;\label{alg:cover:f3E}
}
\Other{\label{alg:cover:fOS}
	\If{$\langle A_l, B_l\rangle$ \rm is small}{
		\textsc{CoverConcept}($\langle A_l, B_l\rangle, l$)\;
		$potential[l] \leftarrow 0$\;
	}
	\Else{
		\textsc{CoverIncremental}($\langle A_l, B_l\rangle, l, bestCoverage$)\;
	}
	\Return{$covers[l]$}\;\label{alg:cover:fOE}
}
}
\end{algorithm}

\subsubsection{Uncovering concepts}

Process of uncovering ones is very similar to that of GreCon2, see Algorithm~\ref{alg:uncov}. For each one covered by the given formal concept a list of concepts is obtained (line~\ref{alg:uncover:foreach}). Then, coverage of each such concept is decreased. Unlike GreCon2, whenever coverage of a concept and its potential to cover ones reaches zero, its slot is set to be free and may be used by some other concept (lines \ref{alg:uncover:free1} and~\ref{alg:uncover:free2}).

\begin{algorithm}[ht]
\small
\caption{\textsc{Uncover($\langle A_l, B_l\rangle$)}}
\KwIn{Formal concept $\langle A_l, B_l\rangle$ that was selected as a factor concept.}
\label{alg:uncov}
\ForEach{$i \in A_l$}{
	\ForEach{$j \in B_l$}{
		\ForEach{$k \in cell[i \cdot n + j]$}{\label{alg:uncover:foreach}
			$covers[k] \leftarrow covers[k] -1$\;
			\If{$(covers[k] + potential[l]) = 0$} {\label{alg:uncover:free1}
				$concepts[k] \leftarrow$ {\bf empty}\;\label{alg:uncover:free2}
			}
		}
		delete list $cell[i \cdot n + j]$\;
	}
}
\end{algorithm}

\begin{remark}
The \textsc{Uncover} procedure works only with values in the $cells$ and $covers$ arrays that were initialized in the \textsc{Cover} procedure. It does not affect values in the $potential$ array. Therefore, if coverage of some partially processed concept (i.e., concept with $potential[l] > 0$) is adjusted, only a value of $covers[l]$ is adjusted, hence $covers[l]$ and $potential[l]$ contain sound and expected values.
\end{remark}

\section{Implementation and Evaluation}
\label{sec:evaluation}

We implemented the proposed algorithm in order to evaluate it and provide a comparison with the state-of-the-art algorithms. In what follows, we discuss the implementation of GreCon3 and provide its experimental evaluation, including comparison with GreCon2 and GreCon3.\footnote{The original GreCon algorithm is not included since GreCon2 outperforms it.} 


\subsection{Implementation notes}

We implemented all algorithms in the Java programming language with the same optimization level to make the comparison as fair as possible. Furthermore, for the sake of fairness, all algorithms use the same data structures where possible.
The implementation of GreCon2 follows its original description from~\cite{TrVy22:KBS}, implementation of the GreConD algorithm incorporates optimizations discussed in~\cite{KrTr19:ParaGreCond}.

\subsection{Datasets}

In the experimental evaluation, we used standard real-world benchmark datasets commonly employed in BMF research. Namely, we used: advertisement and mushroom from the UCI ML Dataset Repository~\cite{uci};
apj, americas\_large, americas\_small, and customer from~\cite{exact};
DNA from~\cite{MyDCNAPoHN}; nfs from the UCI KDD Repository\footnote{\url{https://kdd.ics.uci.edu/}};
and T10I4D100K from the FIMI Dataset Repository.\footnote{\url{http://fimi.uantwerpen.be/data/}} In addition, we included several datasets binarized from data from~\cite{uci}. Specifically, we used ord5bike\_day, nom20magic, inter6shuttle and inter10crx.\footnote{All of them are avalable at \url{https://phoenix.inf.upol.cz/~konecnja/fcalad/}.}

The characteristics of the datasets are shown in Table~\ref{tab:datasets}. Specifically, the number of objects, the number of attributes, the density of nonzero entries in the data (in percents), and the number of concepts.

\begin{table}[ht!]
\small
\caption{Datasets and their characteristics.}\label{tab:datasets}
\centering
\begin{tabular}{lrrrr}
	\toprule
	dataset         &   objects & attributes & density (\%) &   concepts \\ \midrule
	advertisement   &  $3,279 $ &    $ 1557$ &         0.88 &      9,192 \\
	americas\_large &  $10,127$ &    $3,485$ &        52.50 &     36,992 \\
	americas\_small &  $3,477 $ &   $ 1,587$ &         1.91 &      2,764 \\
	apj             &  $2,044 $ &   $ 1,164$ &         0.29 &        798 \\
	customer        & $10,961 $ &     $ 277$ &         1.50 &     47,848 \\
	dna             &  $4,590 $ &     $ 392$ &         1.47 &      4,483 \\
	Mushroom        &  $8,124 $ &     $ 119$ &        17.65 &    22,1525 \\
	nfs             &    12,841 &      4,894 &        89.82 & 24,303,286 \\
	T10I4D100K      &    100,000       &1,000           &             1.01 &    2,347,376        \\
	ord5bike\_day   &       731 &         58 &        35.18 &     81,277 \\
	nom20magic      &    19,020 &        202 &         5.45 &  1,376,212 \\
	nom15magic      &    19,020 &        152 &         7.24 &  1,149,717 \\
	inter6shuttle   &    43,500 &        106 &        43.44 &    381,636 \\
	inter10crx      &       653 &        139 &        44.10 & 10,199,818 \\ \bottomrule
\end{tabular}
\end{table}

\subsection{Running times comparison}

We compared the running times of {GreConD}, {GreCon2}, and {GreCon3}. All experiments were performed on otherwise idle computer equipped with AMD Ryzen Threadripper 3960X (24 cores $\times$ 2 threads), 188 GB RAM, running Ubuntu Linux and Java 21. For the purposes of our experiments the heap size of Java Virtual Machine was set to 100 GB. We run each algorithm 100 times on a particular dataset. The minimal time is reported in Table~\ref{tab:running-times1} and \ref{tab:running-times2}. In the case of \textsc{GreCon2} and \textsc{GreCon3}, the time required for the computation of all formal concepts is included. To obtain all formal concepts the 3,4-CbO~\cite{KoKr:50shadesOfCbO} algorithm is used.

\begin{table}[ht!]
\small
\caption{Comparison of running times of {GreConD}, {GreCon2} and {GreCon3} in miliseconds.}\label{tab:running-times1}

\centering
\begin{tabular}{llrrrrrr}
	\toprule
	dataset         & algorithm &  0.75 &   0.8 &  0.85 &    0.9 &  0.95 &      1 \\ \midrule
	advertisement   &   GreConD &   118 &   136 &   159 &    198 &   287 &    540 \\
	                &   GreCon2 &   128 &   121 &   120 &    129 &   127 &    134 \\
	                &   GreCon3 &   138 &   142 &   134 &    116 &   124 &    118 \\ \midrule
	americas\_large &   GreConD &   526 & 1,267 & 1,801 &  2,823 & 3,164 & 14,656 \\
	                &   GreCon2 & 2,146 & 2,008 & 2,014 &  2,003 & 1,932 &  1,979 \\
	                &   GreCon3 & 1,158 & 1,308 & 1,198 &  1,244 & 1,278 &  1,324 \\ \midrule
	americas\_small &   GreConD &    43 &    76 &    70 &    114 &   146 &    727 \\
	                &   GreCon2 &   135 &   140 &   138 &    140 &   140 &    120 \\
	                &   GreCon3 &    88 &    90 &    92 &     93 &    96 &     84 \\ \midrule
	apj             &   GreConD &    50 &    58 &    71 &     84 &   100 &    133 \\
	                &   GreCon2 &    23 &    22 &    24 &     20 &    21 &     20 \\
	                &   GreCon3 &    20 &    22 &    20 &     20 &    20 &     18 \\ \midrule
	customer        &   GreConD &    42 &    44 &    52 &     60 &    71 &    148 \\
	                &   GreCon2 &   162 &   150 &   154 &    150 &   156 &    142 \\
	                &   GreCon3 &   122 &   114 &   120 &    124 &   122 &    128 \\ \midrule
	dna             &   GreConD &    45 &    52 &    57 &     73 &    98 &    197 \\
	                &   GreCon2 &    26 &    27 &    27 &     26 &    25 &     25 \\
	                &   GreCon3 &    26 &    26 &    26 &     28 &    27 &     28 \\ \midrule
	mushroom        &   GreConD &    42 &    48 &    54 &     66 &    76 &    124 \\
	                &   GreCon2 & 9,306 & 9,556 & 9,262 & 10,252 & 9,439 & 10,304 \\
	                &   GreCon3 & 1,732 & 1,899 & 1,948 &  2,046 & 2,079 &  2,174 \\ \bottomrule
\end{tabular}
\end{table}

\begin{table}[ht!]
\small

\caption{Comparison of running times of {GreConD}, {GreCon2} and {GreCon3} in miliseconds. The ``–'' symbol indicates that the corresponding algorithm was unable to compute the factorization due to insufficient memory.}\label{tab:running-times2}

\centering
\begin{tabular}{llrrrrrr}
	\toprule
dataset         & algorithm &   0.75 &    0.8 &   0.85 &    0.9 &   0.95 &      1 \\ \midrule

	nfs             &   GreConD &   6,460 &   7,372 &   8,203 &  11,048 &  14,355 &  21,178 \\
	                &   GreCon2 & 137,062 & 141,348 & 146,167 & 152,572 & 163,548 & 188,488 \\
	                &   GreCon3 & 112,488 & 112,903 & 113,273 & 115,073 & 113,894 & 121,254 \\ \midrule

	T10I4D100K      &   GreConD &   3,026 &   3,310 &   3,570 &   3,926 &   4,998 &   6,078 \\
	                &   GreCon2 &  10,534 &   9,944 &  10,592 &  10,608 &  10,528 &  10,790 \\
	                &   GreCon3 &   5,350 &   5,402 &   5,450 &   5,416 &   5,558 &   6,036 \\
\midrule

ord5bikes & GreConD & 90 & 101 & 119 & 138 & 166 & 442 \\
 & GreCon2 & 934,334 & 964,158 & 1,038,218 & 1,049,886 & 984,916 & 945,649 \\
 & GreCon3 & 135,037 & 199,301 & 211,154 & 217,068 & 220,219 & 228,786 \\
\midrule

nom20magic & GreConD & 81 & 92 & 105 & 120 & 134 & 218 \\
 & GreCon2 & 5,000 & 4,838 & 4,863 & 4,822 & 4,844 & 4,948 \\
 & GreCon3 & 2,039 & 2,054 & 2,024 & 2,100 & 2,098 & 2,382 \\
\midrule

nom15magic & GreConD & 64 & 70 & 79 & 88 & 105 & 164 \\
& GreCon2 & 5,256 & 5,439 & 5,441 & 5,496 & 5,327 & 5,254 \\
 & GreCon3 & 1,764 & 1,792 & 1,794 & 1,829 & 1,827 & 2,150 \\
\midrule

inter6shuttle & GreConD & 183 & 180 & 208 & 255 & 290 & 957 \\
 & GreCon2 & – & – & – & – & – & – \\
 & GreCon3 & 35,452 & 69,554 & 222,932 & – & – & – \\
\midrule

inter10 & GreConD & 12 & 13 & 16 & 20 & 31 & 90 \\
 & GreCon2 & – & – & – & – & – & – \\
 & GreCon3 & 185,406 & 266,860 & 283,436 & 309,019 & 348,688 & 364,244 \\

\bottomrule
\end{tabular}
\end{table}

\begin{remark}
All evaluated algorithms are implemented in the Java programming language.  From our previous experience, Java is a sufficiently efficient programming language and platform for this kind of algorithm and provides competitive performance to C or C++. On the other hand, it compiles and optimizes programs in a just-in-time manner, thus the first runs of the algorithm may be slower than the subsequent runs. Therefore, the minimal running time is reported.
\end{remark}

We also evaluated the running times in cases where only approximate factorization is required. Columns marked by 0.75, 0.8, \dots, 1 capture times required for obtaining a prescribed coverage of input entries, e.g., 0.95 means that the computed factorization covers 95\% of nonzero entries.

\subsection{Discussion}
As we can see from Table~\ref{tab:running-times1} and \ref{tab:running-times2}, GreCon3 significantly outperforms GreCon2, especially on larger datasets. Further, results indicate that GreCon3 is capable of processing datasets that are infeasible for earlier versions of the GreCon algorithm. We also observe that GreCon3 is more advantageous in scenarios where an exact factorization is not required. This is mainly due to its on-the-fly initialization of the data structure. 

Moreover, we observed (in line with \cite{TrVy22:KBS}) that both GreCon2 and GreCon3 sometimes outperform GreConD, which is currently considered one of the fastest BMF algorithms. This behavior is particularly evident in cases with a larger number of attributes and a smaller number of concepts.

\section{Conclusion and Future Research}
\label{sec:conclusion}
In the paper, we introduced GreCon3, a new algorithm for Boolean matrix factorization that revisits and substantially improves the GreCon family of algorithms. The main contribution lies in a novel, memory-efficient data structure for maintaining coverage information, together with a set of optimizations that reduce both runtime and memory requirements. These optimizations include on-the-fly initialization of candidate concepts, incremental computation of coverage, and specialized handling of the first few factors using 
the set properties of factor concepts. Together, they allow GreCon3 to process significantly larger datasets than GreCon2 while providing a substantial speedup.

The experimental evaluation demonstrates that GreCon3 consistently outperforms GreCon2 and sometimes GreConD. The advantage becomes even more pronounced when approximate decompositions are considered, as GreCon3 avoids unnecessary initialization and focuses only on relevant concepts.

Future research may proceed in several directions.
Even though GreCon3 significantly reduces the memory footprint of the algorithm compared to GreCon and GreCon2, in some scenarios, the algorithm is still limited by the available memory. Therefore, we should focus on more memory-efficient representations of data structures used by the algorithm. Namely, we are to focus on a more compact data structure for formal concepts and a data structure holding information on uncovered ones.  The GreCon3 algorithm does not need to keep all formal concepts in the main memory of a computer. Therefore, it should be possible to design an algorithm that keeps the majority of formal concepts in the secondary memory of a computer (e.g., hard drive) and reads them only when necessary, thus further reduces the memory demands of the algorithm and allows the processing of larger data.



\end{document}